\documentclass{appolb}
\usepackage{graphicx}
\newlength{\fight}
\newcommand{\fgb}[3]
{\begin{figure}[tb]
\resizebox{\fight}{!} {\includegraphics{#1}}
\caption{#2}\label{#3}\end{figure}}
\newcommand\ltdash{\raise-1.8pt\hbox{$\scriptscriptstyle |$}}
\newcommand \beq  {\begin{equation}}
\newcommand \eeq  {\end{equation}}
\newcommand \bea {\begin{eqnarray} }
\newcommand \eea {\end{eqnarray}}

\newcommand\dg{^{\dagger}}

\newlength{\bxwidth}\bxwidth=1.5 truein

\newlength{\figwidth}
\newlength{\shift}
\newcommand{\fg}[3]
{
\begin{figure}[htb]
\vspace*{-0cm}
\[
\includegraphics[width=\figwidth]{#1}
\]
\vspace*{\shift}
\caption{\label{#2}
\small
#3
}
\end{figure}}
\begin{document}
\title{Heavy Electron Quantum Criticality
\thanks{Presented at the Strongly Correlated Electron Systems 
Conference, Krak{\'o}w 2002}%
}
\author{ P.Coleman$^1$ and C. P{\'e}pin$^2$
\address{$^1$
Center for Materials Theory, 
Department of Physics and Astronomy,
Rutgers University,
136 Frelinghausen Road,
Piscataway, NJ 08854-8019, USA
}
\address{$^2$
SPhT,CEA-Saclay, l'Orme des Merisiers,
91191 Gif-sur-Yvette, France.
}
}
\maketitle
%
%
%
\begin{abstract}
Although the concept of a quantum phase transition has been known
since the nineteen seventies,  
their importance as a source of radical
transformation in metallic properties has only recently been
appreciated.  A quantum critical
critical point forms an essential singularity in the phase diagram
of correlated matter. We discus new insights into
the nature of this phenomenon recently gained from experiments in
heavy electron materials. 
\end{abstract}
\vskip0.1in

\section{The Challenge of Quantum Criticality}\label{}

Over the past few years, condensed matter physicists have become
fascinated by the phenomenon of quantum criticality. 
Classical phase transitions at finite
temperature involve the development of an 
an order parameter $\psi $. 
A material that is tuned close to a classical phase transition
senses the imminent change of state as the order parameter develops thermal fluctuations
over larger and larger regions of the sample, ultimately forming a
scale-invariant state of fluctuating order called a 
"critical state".  The understanding of the 
universal nature of the correlations that develop at a classical critical point 
is a triumph of twentieth century physics.\cite{wilson}

The analogous idea of quantum criticality was introduced by John Hertz
during the hey-days of interest in critical
phenomena, but was regarded as an intellectual
curiosity.\cite{hertz} 
Discoveries over the past decade and a half have radically changed
this perspective, revealing
the ability of quantum phase transitions to qualitatively transform
the properties of a material at finite temperatures. For
example, high temperature superconductivity is thought to be born from
a new metallic state that develops at a certain critical doping in
copper-perovskite materials.\cite{loram}  Near a quantum phase transition, a
material enters a weird state of "quantum criticality": a new state of
matter where the wavefunction becomes a fluctuating entangled mixture
of the ordered, and disordered state. The physics that governs this new
quantum state of matter represents a major unsolved 
challenge to our understanding of correlated matter. 

\shift=-0.2cm
\begin{center}
\figwidth=0.8\textwidth \fg{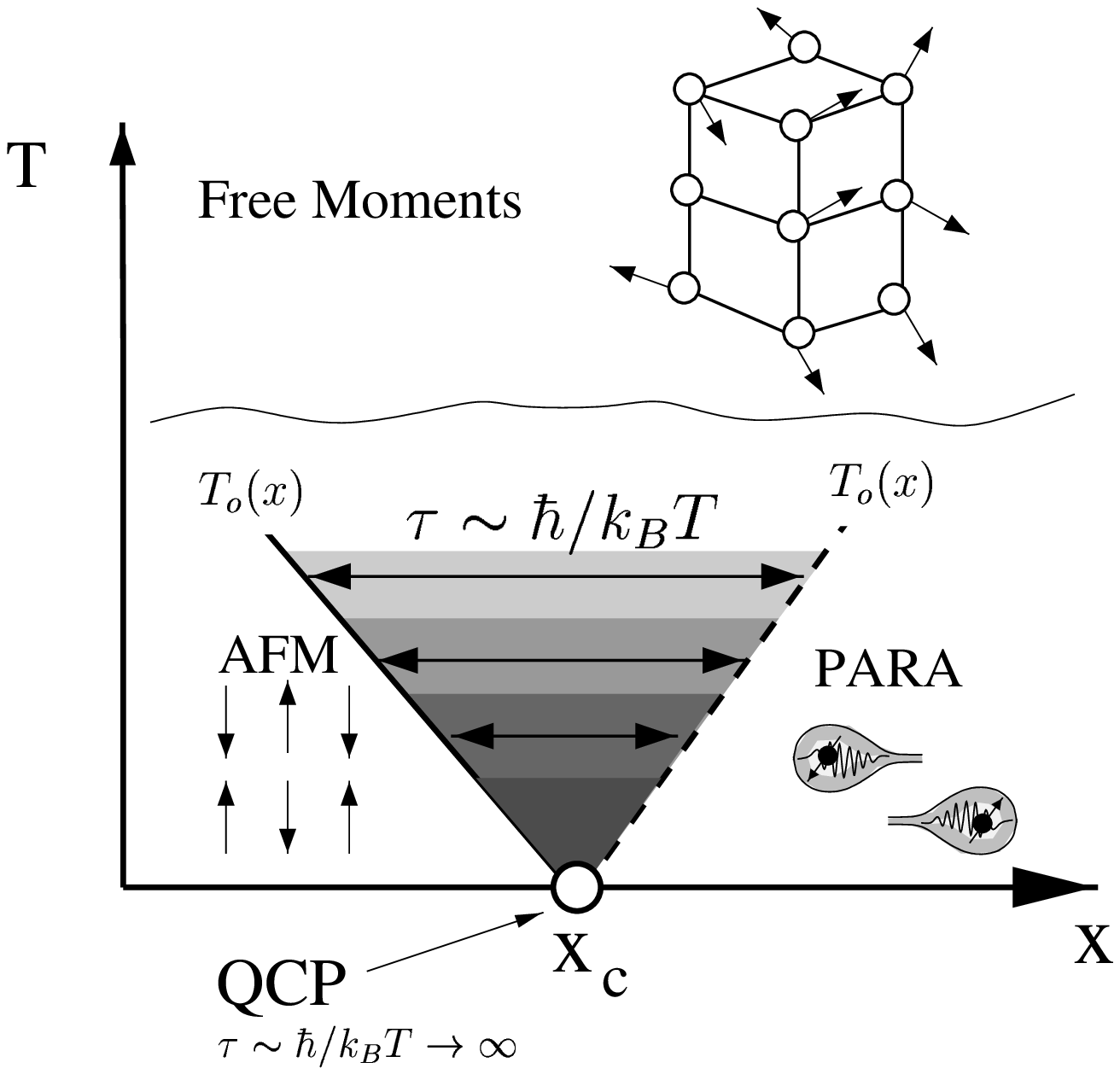}{fig0}{Quantum criticality
in heavy electron systems. High temperature: local moments. For
 $x<x_{c}$ spins become ordered for $T<T_{o} (x)$ forming
an antiferromagnetic Fermi liquid  ; for $x>x_{c}$, 
composite bound-states form between spins and electrons  at $T<T_{0}
(x)$ producing a
heavy Fermi liquid.  ``Non-Fermi liquid behavior'', in which the
characteristic energy
scale is temperature itself, develops in the wedge shaped region between these two
phases. 
} 
\end{center}
\shift=-0.9cm

A quantum critical point (QCP) is a singularity in the phase diagram:
a point $x=x_{c}$ at zero-temperature 
where the
characteristic energy scale $k_{B}T_{o} (x)$ of excitations above the ground-state
goes to zero. (Fig.~1.).\cite{sachdevbook,continentino,moriya,millis,varma2001} 
The QCP 
affects the broad  wedge of phase
diagram where $T> T_{o} (x)$.  
In this region of the material
phase diagram, the critical quantum fluctuations are cut-off by
thermal fluctuations after a correlation time given by the Heisenberg 
uncertainly principle\footnote{Scaling of $\tau\sim \hbar/{k_BT}$ is an example of ``na{\"\i}ve'' scaling and is only expected to occur in 
quantum critical systems that lie below their upper
critical dimension. 
}
\[
\tau \sim \frac{\hbar }{k_{B}T}.
\]
As a material is cooled towards a quantum critical point,
the physics probes the critical quantum fluctuations
on longer and longer timescales. 
Although the ``quantum critical''  region of the phase
diagram where $T>T_{o} (x)$ is  not a strict phase, 
the absence of any scale to the excitations other
than temperature itself qualitatively transforms the properties of the
material in a fashion that we would normally associate with a new
phase of matter.  

Heavy electron materials, offer a unique opportunity to study 
quantum criticality in a metal
where the symmetry and character of the ground-state on either side of
the QCP is unambiguous.  These materials contain
a dense array of local moments derived from rare earth or actinide
atoms, embedded in a conducting host. 
At high temperature they 
display  a Curie-Weiss temperature dependence of the magnetic susceptibility
$
\chi (T) \sim \frac{1}{T}
$
that is the hallmark of local moment metals. 
Depending on the exact conditions of the material, these local moments
can order, forming an antiferromagnetically ordered  metal,  
or they form composite bound-states with the 
the surrounding electrons,  giving rise to 
a highly renormalized Landau Fermi liquid.\cite{landau}

There is a growing list of heavy electron materials that 
can be tuned into the quantum critical
point,  by
alloying, such as $CeCu_{6-x}Au_x$\cite{hvl}, 
through 
the direct application of pressure, as in the 
case of $CeIn_3$\cite{mathur} and $CePd_2Si_2$\cite{grosche} 
or via the application
of a magnetic field, as in the case of $YbRh_2Si_2$.\cite{gegenwartce,gegenwartfieldtuned}  
The recently discovered ``1-1-5'' materials\cite{sarrao1,sarrao2,sarrao3} 
also appear to lie remarkably close to quantum criticality, with
examples of chemically-, pressure-($CeRhIn_{5}$\cite{sarrao1}) and field- tuned
quantum criticality  ($CeCoIn_{5}$.\cite{sarrao2,sarrao3}

\section{Key Properties.}\label{}

In the ground-state near a quantum critical point, 
heavy electron materials display a linear specific heat $C_{V}=\gamma
T$,  and a
quadratic temperature dependence of the resistivity 
$\rho= \rho _{0}+A T^{2}$.
Both of these properties are characteristic of Landau Fermi liquid.
As the QCP is approached, both $\gamma$ and $A$ appear to diverge,
indicating a divergence in the effective mass at the QCP. 

Some of the key properties at the 
QCP are :
\begin{itemize}
\item a divergent specific heat coefficient $\gamma (T)= C_{V}/T$ 
\cite{steglichmass,aoki,schroeder}, which often displays 
a logarithmic temperature dependence\cite{sereni}
 \bea
 \gamma(T) = \gamma_0 \log[ \frac{T_0}{T} ] \ .
 \eea

\item  a quasi-linear temperature dependence of the resistivity 
.\cite{grosche,gegenwartce,julian}
 \beq \rho \propto T^{1+ \epsilon} , \eeq with $\epsilon$ in the
 range of $0-0.6$. 
Many compounds, such as
 YbRh$_2$Si$_2$~\cite{trovarelli} and
 CeCu$_{6-x}$Au$_x$~\cite{schroeder} and CeCoIn$_5$\cite{sarrao1,sarrao2}
exhibit a perfectly linear resistivity, 
 reminiscent of the cuprate perovskites.

 \item anomalous exponents in the spin susceptibility,
$\chi^{-1}(T) -\chi_0^{-1} \sim T^a $, 
 with $a <1$ for CeCu$_{5.9}$Au$_{0.1}$,
 YbRh$_2$(Si$_{1-x}$Ge$_x$)$_2$ ($x=0.05$) and
CeNi$_2$Ge$_2$.\cite{grosche}
In CeCu$_{6-x}$Au$_x$,\cite{schroeder}
neutron scattering measurements 
\cite{schroeder} reveal $\omega/T$\cite{aronson} 
scaling in the dynamic
spin susceptibility
 \begin{equation}\label{lab1}
 \chi^{-1}( {\bf q},\omega ) = f({\bf q}) + ( i \omega + T )^a.
 \end{equation}
where $ f({\bf q})\rightarrow 0$ at the ordering wave vector(s). \end{itemize}

The appearance of temperature as the only energy scale in the critical
spin fluctuations
with a non-trivial exponent $a<1$, is an example of ``na{\"\i}ve scaling'',
where the boundary condition (in this case, the periodicity of the
fields over the imaginary time $\tau \in ( 0, \hbar /k_{B }T)$) determines the correlation
time. 
This is 
a hallmark of a system where the critical modes lie beneath their
upper critical dimension.\cite{ourreview}
The $q-$independence of damping in the critical spin fluctuations
suggests a local element to the underlying physics, and 
has stimulated efforts to develop a ``locally quantum-critical'' theory
of the heavy electron QCP.\cite{qmsi}

Recently, it has become possible to 
examine  the evolution
of the Fermi liquid properties at asymptotically low temperatures in
the approach to a quantum critical point. Particularly interesting
insights have been obtained from the material $YbRh_{2}Si_{2}$. This
material has a $70mK$ Neel temperature. 
By doping this material with Germanium, to form 
$YbRh_{2} (Si_{1-x}Ge_{x})_{2}$, ($x\sim 0.05$), the N{\'e}el temperature
is driven to zero. In this quantum critical state, a tiny magnetic
field  is sufficient to drive the material into  a Fermi liquid
state. These
studies indicate the presence of a single field-tuneable
energy scale in both the specific heat $C_{V}/T$ and the resistivity
$\rho (T)$,  The resistivity shows a field dependent cross-over
between quadratic and linear temperature dependence, whilst the
specific heat shows a field-dependent cross-over between 
a low-temperature upturn of the form $C_{V}/T\sim
1/T^{1/3}$ at $T>>b$ and $C_{V}/T\sim 1/b^{1/3}$, where $b=B-B_{c}$,  in the
field-tuned Fermi liquid. These results  
 can be parameterized in the following form
\begin{eqnarray}\label{}
\frac{d\rho }{dT} &\sim&  f\left(
\frac{T}{T_{o} (b)} 
\right)\cr
\frac{C_{V}}{T} &\sim& \frac{1}{T^{1/3}} \Phi \left(
\frac{T}{T_{o} (b)}  \right)
\end{eqnarray}
where $T_{o} (b)\propto b $, and 
$ f (x)\sim  \min (x,1)$, $\Phi  (x)\sim (\min
(x,1))^{1/3}$. (Fig. 2.)
\shift=-0.2cm
\begin{center}
\figwidth=0.65\textwidth \fg{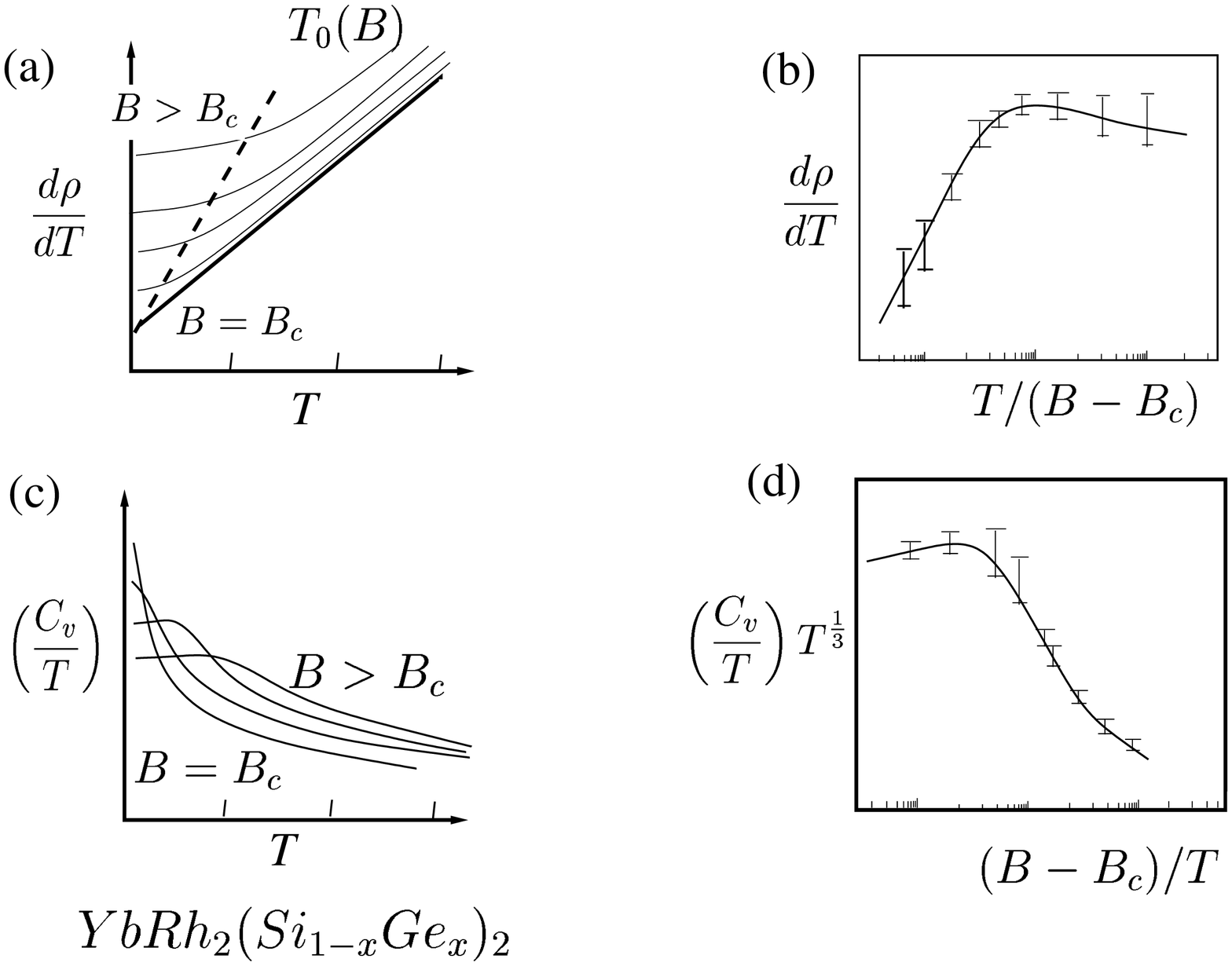}{fig1}{Cartoon illustrating
how the evolution of the resistivity and specific heat in $YbRh_{2} (Si_{1-x}Ge_{x})_{2}$
($x\sim 0.01$) is determined by a single scale $T_{0} (B)\sim
(B-B_{c})$, after \cite{custers}. (a) Linear resistance at criticality develops into
quadratic dependence away from the QCP, (b) scaling of $\frac{d\rho
}{dT}$ (error bars indicate spread of data),  (c) field dependence of
specific heat coefficient and  and 
(d) scaling of the specific heat coefficient $C_{V}/T$. 
} 
\end{center}
\shift=-0.4cm
\vskip -0.2cm
\noindent The existence of a single scale both the thermodynamics and the
transport properties is striking
evidence for the idea that the Fermi temperature goes to zero at a
heavy fermion QCP.   These results place very severe constraints on our
understanding of the physics, as we now discuss. 

\section{Difficulties with the Standard Model}\label{}

The ``standard model'' of heavy fermion quantum criticality, is provided
by the Moriya-Hertz-Millis quantum spin density wave (QSDW) theory.\cite{hertz,moriya,ueda,millis} 
In this approach, critical behavior results from Bragg diffraction of electrons off quantum 
fluctuations in the spin density, described by an interaction of the form
$
H_I  =g\sum_{\vec{q}}\vec M_{\vec q} \cdot \psi\dg_{\vec k - \vec q}
\vec \sigma \psi_{\vec k}
$.
When the fermions
are integrated out of the physics, the effective action for the
slow quantum spin density modes is assumed to be local, and given by 
\bea
\frac{F}{k_{B}T} = \sum_{Q\equiv(\vec q, i \nu_n)} \vert M(Q) \vert^2 \chi^{-1}(Q)
+ \frac{U}{4\!} \int_{0}^{\frac{\hbar }{k_{B}T}} d \tau \int d^dx M(x,\tau)^4
\eea
The inverse susceptibility 
\bea
\chi^{-1}(Q)=\left((\vec q- \vec Q_0)^2 + \xi^{-2} +
\frac{\vert\nu_n\vert}{\Gamma_{\bf Q} }\right)\chi _{0}^{-1}  
\eea
has an Orenstein-Zernicke form, where $\xi$ is the correlation length,
$\vec Q_0$ is the ordering wave-vector and 
the damping term, linear
in frequency $\nu_n$ derives from coupling to the particle-hole excitations
of the Fermi sea. 

Critical fluctuations in this model strongly scatter electrons
on ``hot lines'' around the Fermi surface which are separated by momentum $\vec
Q_0$ (Fig.~3A (iii)). On the 
hot lines, the electron scattering rate $\Gamma_{sc}\propto \max
(\omega, T)$ is linear in energy and temperature, 
and the quasiparticles  masses are driven to infinity.
This ``marginal''  Fermi liquid behavior\cite{mfl} is confined to a
narrow region of width $\delta k \sim \sqrt{T}$ around the hot lines,
and
even at criticality, the remainder of the Fermi surface would form a tranquil Landau
Fermi liquid.  

In the 3D QSDW scenario, 
the spin correlation time $\tau = \Gamma_{Q_{0}}\xi^{2}$ 
so 
time scales as  {\sl $z=2$ } spatial dimensions. 
The effective spatial dimensionality of the
phase space is $D =d +z $, and since $D_{u}=4$ is the
upper-critical dimension of this kind of ``$\phi^{4} $'' field theory, 
na{\"\i}ve  scaling behavior is only expected for 
$d\leq d_{u}=4-z=2$. 
The 3D QSDW model is thus
inconsistent 
with 
\begin{itemize}

\item  $E/T$ scaling in the spin correlations with a non-trivial exponent

\item  A divergence in the specific heat. 

\item The cross-over to a linear resistivity at $T> T_o(x)$. 

\end{itemize}
It is worth noting that the resistivity of
$CeIn_3$ follows a $T^{1.6}$ variation that is said to be consistent
with the 3D scenario.\cite{knebel} However, recent NMR measurements suggest that this
material has a {\sl first order} transition, so that electrons
never feel the full force of quantum criticality.\cite{kawasaki} 

\shift=-0.5cm
\begin{center}
\figwidth=0.8\textwidth \fg{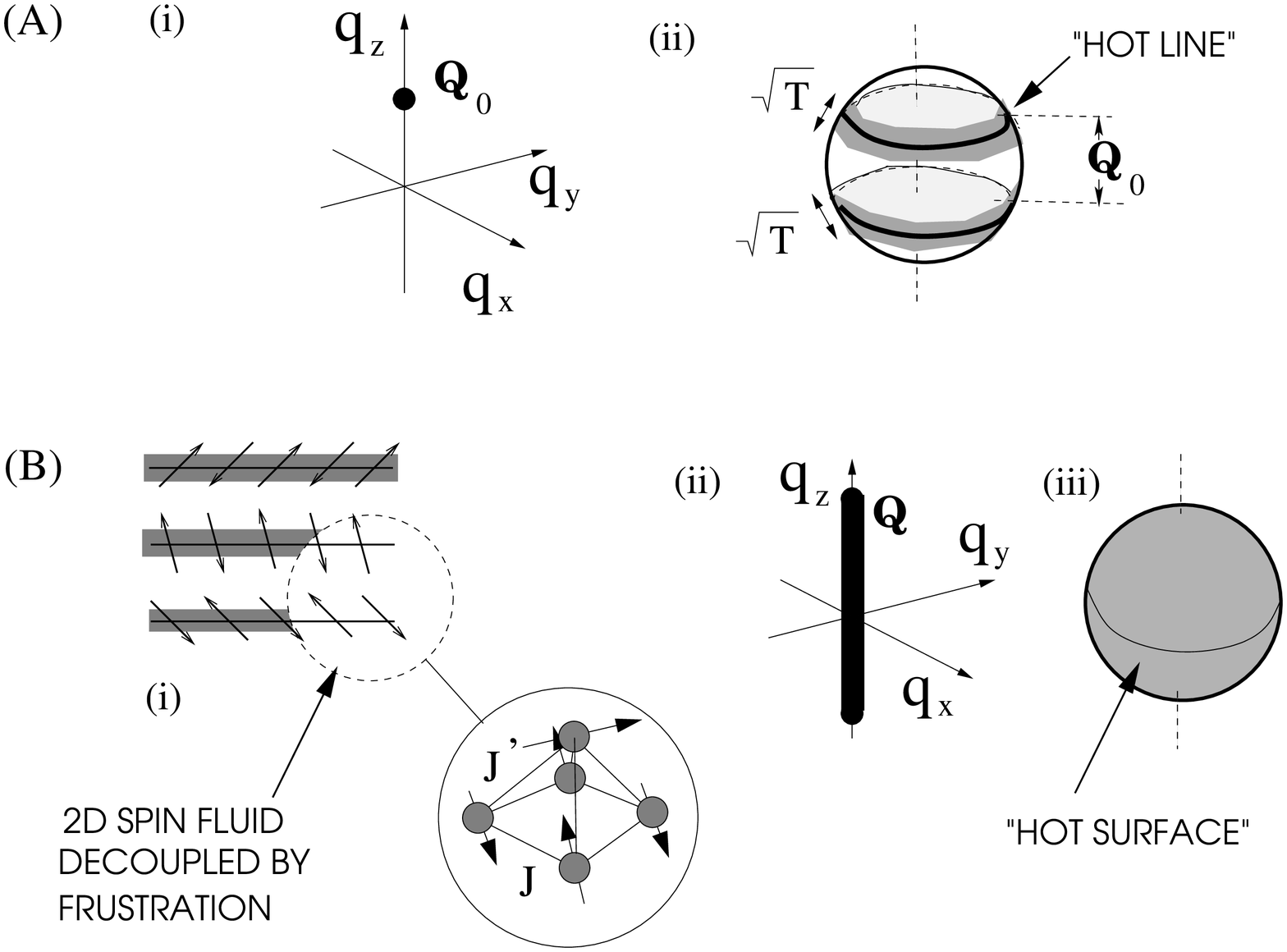}{fig3}{
(A) 3D QSDW scenario in which (i) the critical fluctuations focus around a
point in momentum space giving rise to (ii) hot lines around the Fermi
surface. (B) 2D QSDW scenario, in which (i) frustration leads to layers of decoupled
spin  fluid,  (ii) rods of critical scattering in momentum space and
(iii) non Fermi liquid behavior across the entire Fermi surface. 
} 
\end{center}
\shift=-0.9cm
\vskip -1cm
\section{Is the spin fluid two-dimensional?}

The failure of the 3D QSDW scenario has stimulated the proposal that 
magnetic frustration causes 
the spins to decouple into layers of 
independent two dimensional spin fluids\cite{mathur,rosch} (Fig. 3b). This scenario does predict a logarithmic
divergence in the specific heat of the form \bea \frac{C_v}{T} \sim
\ln \left ( \frac{T_{sf}}{\hbox{max} (T_{3D},T)} \right) \eea where
$T_{sf}$ is the characteristic scale of spin fluctuations and $T_{3D}$
is the scale at which the planes become coupled.
Furthermore, the critical spin fluctuations are 
then critical along ``rods'' in momentum space, \footnote{In quantum critical $CeCu_{6-x}Au_x$ ($x=0.1$) there is evidence for 
rod-like regions of critical fluctuations 
.\cite{rosch,PRL1998Schroeder} }and in this situation large regions of the
Fermi surface become ``hot''. 

Part of the problem with the 2D QSDW, is that we know of no mechanism
to produce such perfectly decoupled 2D spin fluids within three
dimensional metals. 
Even in lattices where frustration decouples spin layers to first order in the
interlayer coupling $J'$, zero point spin
fluctuations couple the spin layers to second
order in the coupling via the mechanism called
``order-from-disorder'',\cite{shendar,henley,larkin} 
$T_{3D}\sim (J')^{2}/J$ and
for this reason, it is 
difficult to suppress $T_{3D}$ more than an
order of magnitude smaller than $T_{sf}$ using frustration.  Yet no
such crossover has been observed, 
indeed, 
$YbRh_{2}Si_{2}$, the specific heat diverges faster than
logarithmically at low temperatures.  

Conventional heavy electron materials form Landau Fermi liquids
which are characterized by local scattering amplitudes. One of the
consequences of this local scattering, is the constancy of the called ``Kadowaki
Woods'' ratio $K=A/\gamma^{2}$ between the quadratic temperature coefficient of the 
resistivity and the square of the specific heat coefficient.\cite{kadowaki} 
This is not expected in the 2D QSDW picture, which will
produce strongly momentum dependent scattering. 
Experimentally, the quadratic $A$ coefficient of the resistivity
diverges in the approach to quantum criticality. From the 
scaling results on field-tuned criticality in $YbRh_2Si_2$  mentioned above, 
$A\sim \frac{1}{T_o(b)}\sim \frac{1}{b}$ ($B=B-B_c$).
Such behavior can be obtained in a two dimensional spin fluid
model in which the inverse squared correlation length is assumed
to be proportional to $b$, $\xi^{-2}\propto b$.\cite{ueda,paulkotliar}  The same model
predicts a weak dependence of the linear specific heat on magnetic
field $\gamma_{th}\propto Log (1/b)$, 
so that 
\[
K_{th}=\left. 
\frac{A}{\gamma^{2}}\right|_{th}\sim \frac{1}{b Log^{2} (b)}.
\]
Early experiments by Gegenwart et al.\cite{gegenwartfieldtuned} suggested that the Kadawaki
Woods ratio is independent of field. 
More extensive scaling results at lower
fields and temperatures \cite{custers}
indicate that $\gamma\sim \frac{1}{b^{1/3}}$ 
so that
\[
K_{exp}=\left.\frac{A}{\gamma^{2}}\right\vert _{exp}
\sim \frac{1}{b^{1/3}}.
\]
This  weak field dependence of the Kadowaki Woods ratio 
indicates that the 
scattering amplitudes in the Fermi liquid do not develop a strong
momentum dependence in the approach to the QCP, arguing against
the exchange soft magnetic fluctuations in  a 2D spin fluid
as the predominant origin of the scattering. 

\section{The Search for New Mean Field Theories}

Traditionally, theories of critical fluctuations 
are built upon an underlying
mean-field theory, which becomes exact above the upper critical
dimension. 
The spin density wave scenario is a consequence of
examining fluctuations about the Stoner and Slater mean-field theory
for itinerant
magnetism.  The failure of this starting point may indicate that we
should search for a new kind of mean-field theory. 
Two ideas have been recently explored: 
\begin{itemize} 

\item {\bf Local Spin criticality.}  The momentum independence
of the spin damping at the QCP 
point\cite{schroeder} has led to the suggestion that the
spin correlations are critical in time, yet 
spatially local\cite{sachdev,siandsmith,sengupta}  permitting their
treatment via the ``extended dynamical mean field theory'' (EDMFT). 
This is a bold  departure from the Wilson- Kadanoff approach to criticality,
for ultimately- only one dimension- time- is active in the critical fluctuations.

\item {\bf Traditional RG approach on a new Lagrangian. }
If we embrace a Wilson-Kadanoff 
approach to quantum criticality,  
then we must seek a new Lagrangian description to 
of magnetism, and the way it couples  to the Fermi Liquid.  
One idea here, is that at the quantum critical point, the heavy
electron breaks-up into its spin and charge components.\cite{questions2}   

\end{itemize}

The momentum-independent scaling term in the inverse dynamic
susceptibility (6)
certainly does suggests that the critical behavior associated with the heavy fermion QCP
contains some kind of {\sl local} critical
excitation.\cite{schroeder} 
Si, Rabello, Ingersent and Smith\cite{qmsi} et al have pursued  the idea that the locally critical
degree of freedom is spin itself.  In their picture, in order that the
characteristic energy scale of local spin fluctuations goes to zero
at the QCP, 
there must be a 
divergent 
{\it local } spin susceptibility
$
\chi_{loc}=\sum_{\vec{q}}\chi (\vec{q},\omega)\vert_{ \omega=0}.
$
The phenomenological form (\ref{lab1}) appears naturally as part of
the EDMFT scheme adopted by Si et al, and by using this form to 
compute the local spin susceptibility, 
\begin{equation}\label{}
\chi _{loc} (T)\sim \int d^{d }q \frac{1}{({\bf q}-{\bf  Q})^{2} +
T ^{\alpha
}}\sim T^{(d-2)\alpha /2}
\end{equation}
Si  et al. conclude that if a divergence of the local spin response 
requires  a two dimension spin fluid. 
They find, based on this assumption, 
that it is possible to reproduce
the anomalous frequency dependence seen in neutron scattering.
\cite{qmsi,grempel}

This intriguing proposal for heavy electron quantum criticality 
still has some important technical hurdles to clear. 
In one interesting development, Pankov, Kotliar and Motome recently reported that the finite
temperature solutions to the EDMFT 
give rise to a first order phase transition
between the antiferromagnetic and paramagnetic phases.\cite{pankov}  
The transition might become second order at zero temperature,
but it is not clear how any scenario with a first order line 
can be simply reconciled
with the finite temperature scaling behavior and the ``fan'' of quantum criticality observed 
in the vicinity of a heavy fermion QCP (Fig. 4.)

\fgb{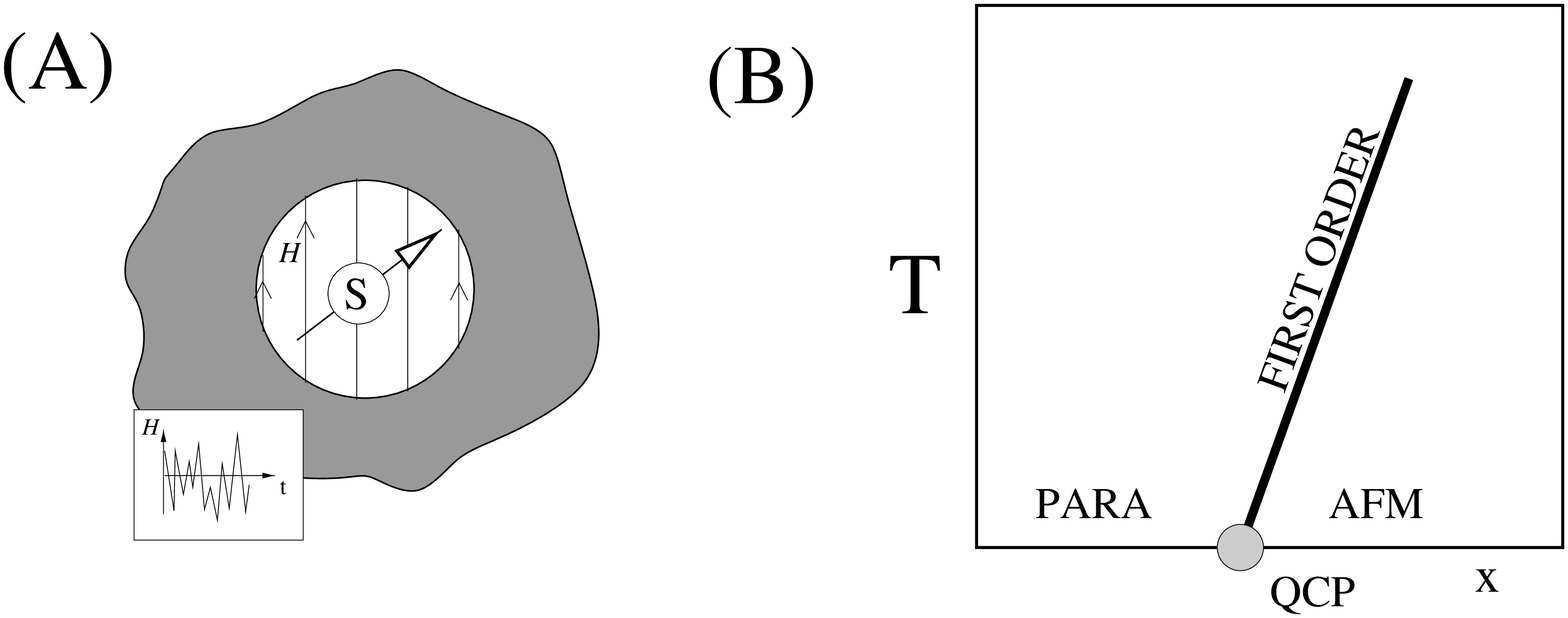}
{In the extended dynamical mean-field theory description of 
local quantum
critical theory, each spin behaves as a local moment in a fluctuating
Weiss field. 
Recent work\cite{pankov} indicates 
that the phase transition predicted by this approach may be 
first order at finite temperature, with a possible QCP at zero temperature
as shown in (b). 
}{fig5}

\section{A new Lagrangian for the emergence of magnetism?}

Another alternative, is that  the 
heavy fermion quantum criticality is 
a a simply three-dimensional phenomenon. In this  
case we need to begin a search for a new class of
critical Lagrangian with an upper critical spatial dimension
$d_{u}>3$.\cite{susy} 
There are a number of elements that might be expected in such a theory: 
\begin{itemize}

\item [$\diamond $]
First- to produce a qualitative departure from conventional
spin fluctuation theory, we should in all probability  seek a new description of the
coupling between the magnetic modes and the heavy electrons. 

\item [$\diamond$]Second-
there is a suspicion that in order to obtain a break-down of the
quasiparticles over the whole Fermi surface, some aspect of the
quantum criticality should be local. 
\end{itemize}

One idea that the current authors have explored, is the notion that
the critical magnetic modes in a heavy fermion system and their
coupling to the Fermi fluid may be 
spinorial in character. 
We know, from various lines of reasoning that 
in a Kondo lattice the Luttinger sum rule,\cite{luttinger} 
\cite{martin82,oshikawa00} governing the Fermi surface volume ${\cal V}_{FS}$
``
counts'' both the electron density $n_{e}$
\underline{and} the number of 
the number of local moments per unit cell
$n_{{s}}$:
 \begin{equation}\label{}
2\frac{{\cal V}_{FS}}{(2\pi)^{3}}= n_{e} + n_{s}.
\end{equation}
The appearance of the spin density in the Luttinger sum rule
reflects the composite nature of the heavy quasiparticles, formed
from 
bound-states between local moments and high energy electron
states. 
Suppose the  spinorial character of the magnetic  degrees of freedom seen in the
paramagnet {\sl also } manifests itself
in the decay modes of the heavy quasiparticles.  This would imply that
at the QCP, the staggered magnetization factorizes 
into a spinorial degree of freedom
$\vec{M} (x) = b\dg(x)\vec{\sigma }b (x) $, where $b$ is a 
two-component bosonic spinor.  
``Spinorial magnetism''
affords a direct coupling between the magnetic spinor $b_{\alpha }$
and the heavy electron quasi-particle fields $\psi_{{\vec{k}}\alpha } $ via an inner product, 
over the spin indices
\begin{equation}\label{spinor}
L_{F-M}^{(2)} = g \sum _{{\bf k}, {\bf q}}[ \phi _{\bf q} \dg 
b\dg _{{\bf k}-{\bf
q}\sigma}
\psi  _{{\bf
k}\sigma }
  +\hbox{H.c}], 
\end{equation}
where conservation of exchange statistics obliges us to 
introduce
of a spinless charge $e$ fermion $\phi $. 
This 
would imply that the composite heavy electron  decays
into a neutral ``spinon''and a spinless charge e fermion 
$e^{-}_{\sigma }\rightleftharpoons s_{\sigma } + \phi ^{-}$.

\shift=-0.5cm
\figwidth=\columnwidth \fg{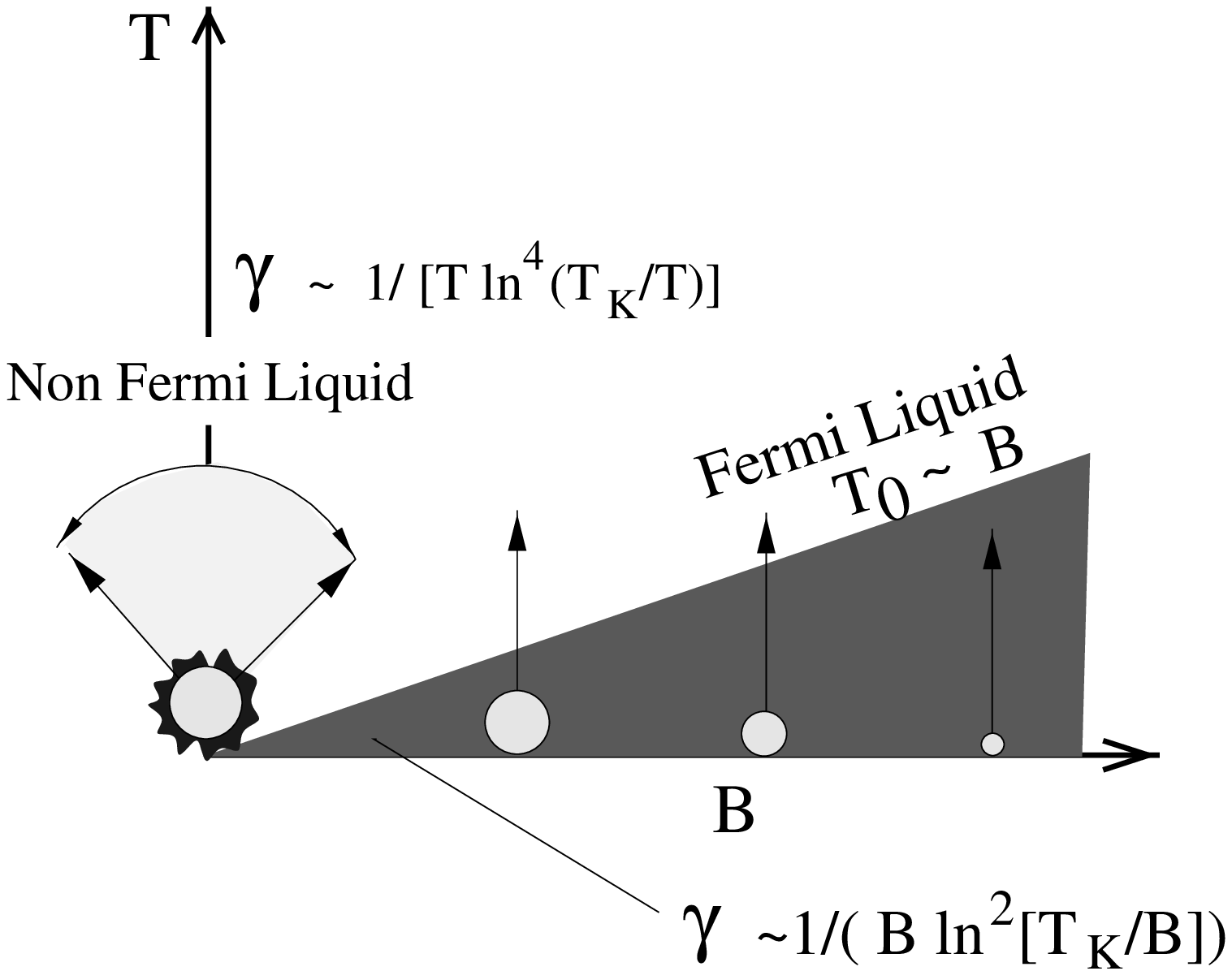}{fig0}{Schematic phase diagram
of the underscreened Kondo model. In a finite field, the ground-state
is a Fermi liquid with a field-tuned Fermi energy. In zero field,
a residual ferromagnetic coupling between the electron sea 
and the untethered moment leads to a break-down of Fermi liquid
behavior and a divergence of $\gamma= C_{V}/T$ with temperature. 
} 
This line of reasoning leads suggests that 
the break-up of the heavy fermion QCP may involve {\sl spin-charge separation}. 
In the antiferromagnet, the magnetic spinors will condense, and the 
$\phi$ fermion will propagate coherently. 
At the QCP, the vanishing of the ordered magnetic moment will mean
that all coherent
motion of this object will cease. 
In such a
scenario, it is then a  ``locally critical''  fermion rather than 
spin that drives the non-Fermi liquid behavior.

Additional 
support for this line of reasoning comes from a quite unexpected
direction:
from the re-examination of a 
venerable
model of magnetism, the ``underscreened Kondo
model'' (UKM).   The underscreened Kondo effect, 
whereby a spin is partially quenched 
from spin $S$ to $S^{*}=S-\frac{1}{2}$. 
occurs in an impurity model when the number of screening channels is
insufficient  to quench the local moment. In impurity models, this
only arises when $S>1/2$. In the Kondo lattice,  underscreening 
may be an intrinsic feature of the 
quantum critical point for $S=1/2$. The Curie-like power-law dependence
$\chi^{-1}(T) -\chi_0^{-1} \sim T^a $, 
of the spin
susceptibility\cite{schroeder,gegenwartfieldtuned} seen at criticality might indeed be interpreted as
circumstantial evidence for the existence of partially quenched  moments
at criticality. 

The UKM  model is written $H=H_{o}+H_{I}$, where $H_{0}$ describes the
conduction sea and 
\begin{eqnarray}\label{start}
H_{I}= 
J \vec S\cdot \psi \dg _{\alpha} \vec \sigma_{\alpha \beta} 
\psi _{\beta}
.
\end{eqnarray}
where $S$ denotes a spin $S>\frac{1}{2}$ and 
$\psi\dg _{\alpha }= \sum_{k}c\dg_{k\alpha }$ creates a conduction
electron at the impurity site. 

In recent work, we have found that the
essential physics of the 
the UKM
is 
captured by a Schwinger boson representation of the local
moments.\cite{boskon} In an unexpected surprise, we have also found that the model
exhibits a unique kind of field-tuned criticality, forming a tunable Fermi
liquid in a magnetic field, but a non-Fermi liquid at $B=0$.
In this approach, 
the  interaction between magnetism
and the Fermi fluid in the UKM 
takes the form 
\begin{equation}\label{}
H_{I}= J \sum_{\alpha ,\beta }\psi \dg_{\alpha }b_{\alpha}
b\dg _{\beta }\psi _{\beta} 
\rightarrow 
\left[
\bar \phi\  
b\dg _{\sigma }\psi _{\sigma } 
+
\psi \dg_{\sigma }b_{\sigma }
\phi 
\right]
-\frac{1}{J}\bar \phi \phi 
\end{equation}
where $\phi $ is a Grassman field. The Gaussian fluctuations of this
field describe a fermionic resonance which couples to the conduction
sea. 
The form of this coupling is suggestively close to the 
phenomenological form (\ref{spinor}) discussed above. 

In a magnetic field $B$, the Schwinger
boson condenses, $\langle b_{\sigma }\rangle
= \sqrt{2M}\delta _{\sigma \uparrow}$, where $M$ is the magnetization, so
that
\begin{equation}\label{}
H_{I}\rightarrow 
\sqrt{2M}\left[
\bar \phi\ \psi _{\uparrow } + {\rm  H. C. } \right] + \hbox{fluctuations}
\end{equation}
giving rise to a resonance in the Fermi sea. What is unexpected about
this resonance, is that its characteristic weight $Z$ or wavefunction renormalization
scales with the magnetic field $B$, $Z\propto B$, giving rise to 
a resonance with a characteristic
energy scale $T_{o} (B)\propto B$.  As the field is reduced to zero,
so the width of the resonance narrows and the linear specific heat can
be shown to diverge as
\[
\gamma
\sim 
\frac{1}{B \ln ^{2} \left(\frac{B}{T_{K}} \right)}.
\]
At zero field the specific heat actually develops a divergence
\[
\gamma
\sim 
\frac{1}{T \ln ^{4} \left(\frac{T_{K}}{T} \right)},
\]
which is reminiscent of the low-temperature upturn in the specific
heat seen in $YbRh_{2} Si_{2}$.
This field-tunability of the Fermi temperature curiously went
un-noticed in 
in the Bethe Ansatz solutions of this model for two decades.\cite{rajan82}
In the new context  it is fascinating because it provides  a
concrete example of a system of field-tuned criticality in a model
where the coupling between the magnetism and the Fermi sea exhibits
an explicit spinorial character. 
One of the
open questions about this model, is whether the temperature dependent 
inelastic scattering it gives rise to will mimic
the a cross-over between quadratic and $T-$ linear 
scattering behavior around $T\sim B$ seen in real heavy electron
systems.  

Finally, we should note that if 
the transition between the antiferromagnet and the paramagnet involves
the formation (or destruction) of new kinds of fermionic resonance at the Fermi
surface, then  the geometry of the Fermi surface will change 
far radically at the heavy electron QCP.
This kind of behavior is expected to give
rise to discontinuities in the Hall conductivity and the extrapolated de Haas van Alphen
frequencies at the QCP.   This is clearly an area where we could
benefit immensely from further  experimental study. 

\section{Summary}
We have reviewed the basic physics of heavy electron
quantum criticality.  The various properties of the antiferromagnetic heavy electron
quantum critical point, most notably the observation  of $E/T$ scaling
and the appearance of a single scale $T_{0} (x)$governing the cross-over from
Fermi liquid, to non-Fermi liquid behavior in both the resistivity
and the thermodynamics,  suggest the existence of a new universality class
of critical electronic behavior that lies beyond the reach 
of quantum spin density wave theories of
quantum criticality.  This motivates a
search for a new class of theory for the emergence of magnetism
in heavy electron systems.  One idea, is that the heavy electron
quantum critical point involves spin correlations that are singular
and critical in time, but only weakly correlated in space, but this
leads to the conclusion that non-trivial behavior requires 
a frustrated, quasi-two dimensional spin fluid. 
Alternatively,  heavy electron quantum criticality may be intrinsically
three dimensional in character, but 
involve a kind of spin-charge
decoupling that develops as the spins bound within 
composite heavy electrons emerge 
into ordered magnetism. 
Our theoretical and experimental explorations of this
phenomenon are still very much in their infancy, and it is clear
that much work remains to be done. 
\\

The work described in this project was  supported
under grant NSF-DMR 9983156 (Coleman). 
The authors gratefully acknowledge their discussions with
N. Andrei, J. Custers, P. Gegenwart, I. Paul, F. Steglich and J, Rech for 
discussions related to this work.

\end{document}